# Variation in correlation between prognosis and histologic feature based on biopsy selection

Emily Diller[1,2] and Jason Parker[1]
[1]Indiana University School of Medicine and [2]Purdue University

## Abstract

Glioblastoma multiform carries a dismal prognosis with poor response to gold standard treatment. Innovative data analysis methods have been developed to characterize tumor genomic expression with histologic features. In a clinical setting, biopsy selection methods may be constrained by time and financial burden to the patient. Thus, we investigate the impact biopsy selection has on correlation between prognostic and histologic features in 35 patients with GBM. We compared methods using limited volumes, moderate volumes, and enblock tumor volumes. Additionally, we investigated the impact of random versus strategic methods for limited and moderate volume biopsies. Finally, we compared correlation results by selecting one to five small biopsy. We observed a wide range in correlation significance across selection methods. These findings may aid clinical management of GBM and direct better biopsy selection necessary for the development and deployment of targeted therapies.

## Introduction

Glioblastoma multiform (GBM) is the most common and worst form of glioma in terms of therapeutic response, aggressiveness, and prognosis. The treatment gold standard for GBM includes surgical resection, chemotherapy, and radiotherapy, which is invasive, has unwanted side effects, and can have a large negative impact on quality of life. Even with the efforts of gold standard treatment, tumor recurrence is very high compared to other neoplasms[1]. Thus, efforts have been made to characterize the molecular and genetic profile within GBM so targeted therapeutic methods could be used with a more effective clinical approach[2,3].

Recently, Ivy Glioblastoma Atlas Project (Ivy GAP) initiated a several studies to evaluate genetic expression associated with histologic and clinical features[4,5]. Cantanhede and Oliveira observed that within histologic features there are significant variations between platelet derived growth factor (PDGR) family genes. Additionally, they found significant differences in expression between left and right lobes.  Similarly, Puchalski et al observed that RNA-seq had significant variation across histologic structural features defined as cellular tumor, leading edge, palisading necrosis, and microvascular proliferation. These works demonstrate with statistical confidence a link between histologic structural features and genetic expression.

In a clinical setting, the diagnostic gold standard is medical imaging, preferably magnetic resonance imaging (MRI), followed by biopsy confirmation. Biopsy tissue volume may vary across cases from needle biopsy to biopsy sectioning during resection. In either case, the tissue sampling is likely less than the average volume per patient, 7 cm$^3$, used by Ivy GAP. Since a GBM hallmark is intratumor heterogeneity, we hypothesized that biopsy features would vary greatly between methods.

## Methods

**Data selection.**

Histologic features, clinical and genomic data for 35 of 41 available GBM patients was retrospective obtained from the Ivy GAP repository (Allen Institute for Brain Science. Ivy Glioblastoma Atlas Project. Available from: https://glioblastoma.alleninstitute.org/). The cohort was selected based on the following inclusion criteria: enbloc resection contains at least three sub blocks; and complete prognosis data. Histologic feature data was acquired from Ivy GAP (Allen Institute for Brain Science. Ivy Glioblastoma Atlas Project. Available from: https://glioblastoma.alleninstitute.org/static/download.html). The histologic features defined by Ivy GAP (SOURCE) are: leading edge (LE); hyperplastic blood vessels in leading edge (LEhbv); infiltrating tumor (IT); hyperplastic blood vessels in infiltrating tumor (IThbv); cellular tumor (CT); perinecrotic zone (CTpnz); psuedopalisading cells but no visible necrosis (CTpnn); pseudopalisading cells around necrosis (CTpan); hyperplastic blood vesselsin cellular tumor (CThbv); microvascular proliferation (CTmvp); and necrosis (CTne). Histologic features were normalized to the total H&E tissue area for each slide. Clinical and genomic data was acquired from Ivy GAP (Allen Institute for Brain Science. Ivy Glioblastoma Atlas Project. Available from: https://glioblastoma.alleninstitute.org/static/download.html) and tabulated for the selected cohort.

**Data analysis.**

From the compiled data, we analyzed seven biopsy selection methods in MATLAB R2018b (www.mathworks.com). Three location methods (random, tumor center, and max necrosis) were applied to two volume methods (slide and block), resulting in the following six combinations: (A) randomly select one slide; (B) randomly select one block average; (C) one slide at tumor centroid; (D) one block average at tumor centroid; (E) one slide at necrosis; and (F) one block average at necrosis. We selected slide and block volumes to represent needle biopsy (NB) and surgical biopsy (SB) volumes respectively. The last method was (G) the enbloc average, representing the max tissue volume of data available for a patient. For methods C and E, if the location of interest contained more than one slide, a slide was chosen at random. For methods B, D, and F all the slides from the corresponding block were averaged together. For biopsy number analysis, we randomly selected one to five slides from unique block locations across each tumor. Correlation between un-censored, continuous variables overall survival (OS), progression free survival (PFS) and histologic features were evaluated by Pearson's correlation. Correlations were considered statistically significant if p value<0.10.

## Results

**Clinical and genomic data for cohort.**

Information from the compiled data as summarized in **Table 1** reveals that the cohort has a young age at the time of diagnosis and comparable gender distribution. For gross primary tumor locations, the right frontal lobe has the highest incident, followed by left parietal and mixed right hemisphere. The cohort has high functioning Karnofsky Performance Status (KPS), with nearly half presenting neurologic defect, indicating mildly impaired quality of life and favorable prognosis (SOURCE 10,11,12). However, as common with GBM, the prognostic outcomes for the cohort are poor with median progression free survival (PFS) of 107 days, and median overall survival (OS) of 439 days. For genomic expression, Isocitrate Dehydrogenase 1 (IDH1) mutation is observed in three patients, consistent with the literature (SOURCE).

| Clinical Data | | |
|---|---|---|
| **Gender** | | |
| *Female* | N=17 | 48.57% |
| *Male* | N=18 | 51.43% |
| **Age At Diagnosis (years)** | | |
| *Mean ± standard deviation* | 58.7 ±12.0 | |
| *Median* | 61 | |
| **Karnofsky Performance Status** | | |
| *Mean ± standard deviation* | 88 ±10.5 | |
| *>70* | N=29 | 82.86% |
| *≤70* | N=6 | 17.14% |
| **History of Seizure** | | |
| *Present* | N=10 | 28.57% |
| *Not-present* | N=25 | 71.43% |
| **Neurologic defect** | | |
| *Present* | N=16 | 45.71% |
| *Not-present* | N=19 | 54.29% |
| **Primary Tumor Location** | | |
| *Left frontal* | N=1 | 2.86% |
| *Left Occipital* | N=1 | 2.86% |
| *Left Parietal* | N=6 | 17.14% |
| *Left Temporal* | N=5 | 14.29% |
| *Right Frontal* | N=7 | 20.00% |
| *Right Parietal* | N=4 | 11.43% |
| *Right Temporal* | N=5 | 14.29% |
| *Mixed Right Hemisphere* | N=6 | 17.14% |
| **Progression free survival (days)** | | |
| *Mean ± standard deviation* | 205.5 ±256.8 | |
| *Median* | 107 | |
| **Overall survival (days)** | | |
| *Mean ± standard deviation* | 491.4 ±321.5 | |
| *Median* | 439.5 | |
| **Genomic Data** | | |
| ***PTEN*** | | |
| *Deletion/Loss* | N=18 | 51.43% |
| *Gain* | N=3 | 8.57% |
| *Normal* | N=3 | 8.57% |
| ***IDH1*** | | |
| *Wild-type* | N=30 | 85.71% |
| *Mutated* | N=3 | 8.57% |
| ***1p19q Deletion*** | | |
| *Positive* | N=3 | 8.57% |
| *Negative/Normal* | N=21 | 60.00% |
| ***MGMT PCR*** | | |
| *Methylated* | N=6 | 17.14% |
| *Unmethylated* | N=19 | 54.29% |

*Table 1. Clinical and genomic data for research cohort. PTEN: Phosphatase and Tensin Homolog; IDH1: Isocitrate Dehydrogenase 1; MGMT: O-6-Methylguanine-DNA Methyltransferase; PCR: Polymerase chain reaction.*

**Correlation between overall survival and histologic feature vary by biopsy method.**

Correlation significance between overall survival (OS) and LE, IT, CTpnz, CTpnn, CTmvp, CTne vary by biopsy method as shown in **Figure 1**. The correlation between OS and LE is significant for NB methods taken at random and at max necrosis (p=0.05, and p=0.09 respectively). For IT, correlation with OS is significant for NB method taken from max necrosis (p=0.08). The correlation between CTpnz and OS is significant for NB methods taken at random and from max necrosis (p=0.05 and 0.07 respectively). For CTpnn, the correlation is significant for SB methods taken at random and at tumor centroid (p=0.008, 0.02 respectively), while for NB method it is significant taken from tumor center (p=0.002). Histologic feature CTmvp has significant correlation with OS for SB methods taken at max necrosis and enblock (p=0.06, and 0.006 respectively). For OS and CTne SB at max necrosis has significant correlation (p=0.04). The correlation between OS and NB, SB, and enbloc is shown in Figure 1 A, C, and E respectively.

**Correlation between progression free survival and histologic feature vary by biopsy method.**

The correlation between progression free survival (PFS) and histologic features CTpnn and CTmvp vary based on biopsy method as shown in **Figure 1**. For SB based methods, the correlation between PFS and CTpnn is significant if the biopsy is taken from the center (p=0.007). The correlation between PFS and CTmvp is significant if NB biopsy from the tumor center or SB from random is used (p=0.03, and p=0.035 respectively). The correlation between PFS and NB, SB, and enbloc is shown in Figure 1 B, D, and E respectively.

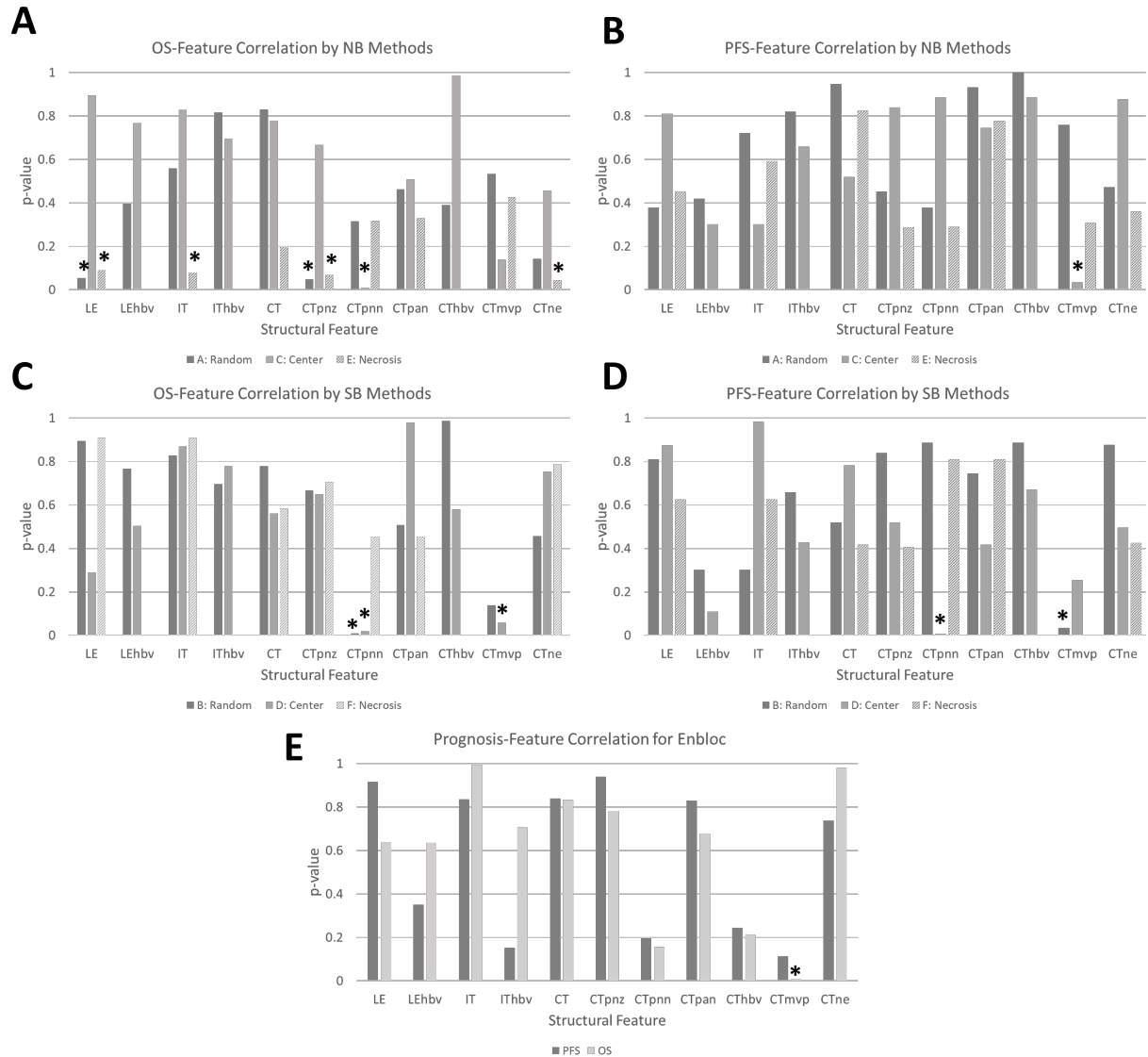

*Figure 1*: *Correlation between prognosis and structural features across biopsy methods. (A) Correlation between overall survival (OS) and histology structural features by limited volume, needle biopsy (NB) like methods. (B) Correlation between progression free survival (PFS) and histology structural features by limited volume, needle biopsy (NB) like methods. (C) Overall survival (OS) correlation to histologic structural features using a larger volume, surgical biopsy (SB) like methods. (D) Progression free survival (PFS) correlation to histologic structural features using a larger volume, surgical biopsy (SB) like methods. (E) Correlation between prognostic factors and histologic structural features using enblock method.*

**Correlation between prognosis and histologic feature vary by the number of biopsies selected.**

For correlation between PFS and histologic feature CTpnz (*p=0.036*) there was significant variation from the number of biopsies. For correlation between OS and histologic features LE

($p=0.026$), LEhbv ($p=0.011$), and CTmvp ($p=0.003$) there was significant variation from the number of biopsies. Variation in correlation between prognosis and histologic features due to the number of biopsies can be observed in **Figure 2**.

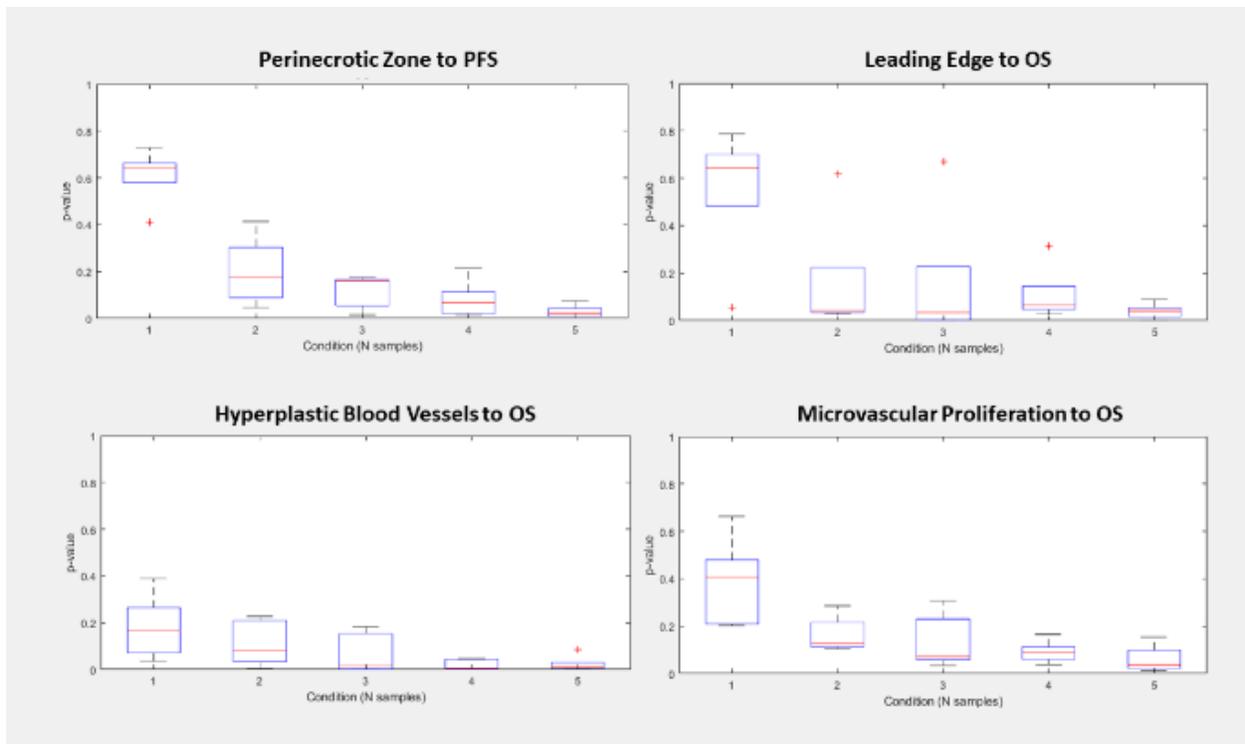

*Figure 2.* Variation in correlation between prognosis and histology based on the number of biopsies taken. PFS: progression free survival; OS: overall survival.

## Discussion

This study presents the impact of biopsy selection method on correlation between prognosis and histologic features. The impact of intra-tumoral heterogeneity can be observed in the correlation range by method. Between needle like biopsy, surgical biopsy, and enblock analysis, correlation with prognosis varies greatly.

GBM has been studied widely, however to our knowledge, this is the first study to look at the effects of sampling technique form in-vivo data. Repositories, such as The Cancer Genome Atlas (TCGA) have clinical, genomic, and biopsy data, however the location of biopsy relative to the tumor is unknown. Additionally, the biopsies volume used to derive histologic features characteristic of GBM[6,7], such as the presence of necrosis or palisading tumor, is unknown. However, multiple studies have linked, or argue such, that a relationship between histologic features and genomic expression and/or imaging features are reliable [8,9].

Based on this analysis, we can observe vast variation in correlation between limited volume biopsies, such as needle biopsy. Between one and three samples there is a wide shift in correlation, consistent with intra-tumoral heterogeneity. Additionally, a plateau is observed in limited volume biopsies between three and five samples, this appears consistent with the observation between random, tumor center, and necrotic core for limited volume biopsy. However, when moving to mid volume or max volume observed, methods B, D, F, and G, few correlations between OS or PFS and histologic features are significant. This appears to represent a "wash-out" like effect, where the heterogeneity across the tumor is no longer observed. Considering the presented data, select limited volume biopsy, such as needle biopsy, across the tumor has the most robust correlation with prognostic factors OS and PFS. Clinically, these methods may be applied to robustly capture histologic features consistent with GBM. Three to five limited volume biopsies taken from the tumor center, max necrosis, and randomly at a minimum of 1 cm separation results in strong correlation to prognostic features.

## References

1. Urbanska, K., Sokolowska, J., Szmidt, M. & Sysa, P. Glioblastoma multiforme - An overview. Wspolczesna Onkol. 18, 307–312 (2014).
2. Omuro, A. & LM, D. Glioblastoma and other malignant gliomas: A clinical review. JAMA 310, 1842–1850 (2013).
3. Vitucci, M., Hayes, D. N. & Miller, C. R. Gene expression profiling of gliomas: merging genomic and histopathological classification for personalised therapy. Br. J. Cancer 104, 545–553 (2011).
4. Puchalski, R., Shah, N., Miller, J., et al. An anatomic transcriptional atlas of human glioblastoma. Science 360, 660-663 (2018).
5. Cantanhede, I., and de Oliveira, J. PDGF Family Expression in Glioblastoma Multiforme: Data Compilation from Ivy Glioblastoma Atlas Project Database. Scientific Reports 7. (2017).
6. Liu, S., Wang, Y., Xu, K., Wang, Z., Fan, X., Zhang, C., Li, S., Qiu, X., and Jiang, T. Relationship between necrotic patterns in glioblastoma and patient survival: fractal dimension and lacunarity analyses using magnetic resonance imaging. Scientific Reports 7. (2017).
7. Raza, S.M., Lang, F.F., Aggarwal, B.B., Fuller, G.N., Wildrick, D.M., and Sawaya, R. Necrosis and Glioblastoma: A Friend or a Foe? A Review and a Hypothesis. Neurosurgery 51, 2–13. (2002).
8. Belden, C.J., Valdes, P.A., Ran, C., Pastel, D.A., Harris, B.T., Fadul, C.E., Israel, M.A., Paulsen, K., and Roberts, D.W. Genetics of Glioblastoma: A Window into Its Imaging and Histopathologic Variability. RadioGraphics 31, 1717–1740. (2011).

9. Barajas, R.F., Hodgson, J.G., Chang, J.S., Vandenberg, S.R., Yeh, R.-F., Parsa, A.T., McDermott, M.W., Berger, M.S., Dillon, W.P., and Cha, S. Glioblastoma Multiforme Regional Genetic and Cellular Expression Patterns: Influence on Anatomic and Physiologic MR Imaging. Radiology 254, 564–576. (2010).